\documentclass[twocolumn,letter]{aastex63}

\usepackage[normalem]{ulem}
\usepackage{footnote}
\usepackage{subfigure}
\usepackage{xspace}
\usepackage[utf8]{inputenc}
\usepackage{multirow,amsmath}
\usepackage{float}
\usepackage{apjfonts}
\usepackage{lipsum}

\newcommand\msun{{M}$_{\odot}$\xspace}                   % solar mass
\newcommand{\BV}{Brunt-V\"{a}is\"{a}l\"{a}}         % brunt

\newcommand{\code}[1]{\texttt{#1}}
\newcommand{\mesa}{\code{MESA} }
\newcommand{\MESA}{\mesa}
\newcommand{\flash}{\code{FLASH} }
\newcommand{\FLASH}{\flash}

\shorttitle{3D Rotating Massive Star Model}
\shortauthors{Fields}

\submitjournal{ApJL}

\begin{document}
\title{\Large  The Three-Dimensional Collapse Of A Rapidly Rotating 16 $M_{\odot}$ Star}

\correspondingauthor{C.~E.~Fields}
\email{carlnotsagan@lanl.gov}

\author[0000-0002-8925-057X]{C.~E.~Fields}
\altaffiliation{Feynman Fellow}
\affiliation{Center for Theoretical Astrophysics, Los Alamos National Laboratory, Los Alamos, NM 87545, USA}
\affiliation{Computer, Computational, and Statistical Sciences Division, Los Alamos National Laboratory, Los Alamos, NM 87545, USA}
\affiliation{X Computational Physics Division, Los Alamos National Laboratory, Los Alamos, NM 87545, USA}

\begin{abstract}
We report on the three-dimensional (3D) hydrodynamic evolution to iron core-collapse of a rapidly rotating 16 \msun star. 
For the first time, we follow the 3D evolution of the angular momentum (AM) distribution in the iron core and convective
shell burning regions for the final 
10 minutes up to and including
gravitational instability and core-collapse.  
In 3D, we find that convective regions show efficient AM transport that leads
to an AM profile that differs in shape and magnitude from \texttt{MESA} within a few shell convective turnover timescales.  
For different progenitor models, 
such as those with tightly coupled Si/O convective shells,  efficient AM transport in 3D simulations could 
lead to a significantly different AM distribution in the stellar interior affecting estimates of the natal neutron star or black hole 
spin.
Our results suggest that 3D AM transport in convective and rotating
shell burning regions are critical components in models of massive stars and could \emph{qualitatively} alter the explosion 
outcome and inferred compact remnant properties. 
\end{abstract} 


\section{Introduction} 
\label{sec:intro}
Massive stars are expected to rotate rapidly at 
their surface during the main sequence (MS)  \citep{fukuda_1982_aa,ramirez_2013_aa,demink_2013_aa}. 
An important aspect of the subsequent evolutionary fate
is the redistribution of angular momentum over its stellar lifetime.  
Work by \citet{heger_2005_aa} investigated the evolution of AM in 
1D stellar evolution models with differential rotation and magnetic fields via the Spruit-Tayler (ST) dynamo 
\citep{spruit_2002_aa}. They found that magnetic torques provided an efficient mechanism for 
transport of AM out of the stellar core to provide remnant spin estimates that were largely in agreement with observations.  
Unfortunately, these models were limited by the treatment 
of convection and different mixing mechanisms which can only be approximated in 1D \citep{jones_2017_aa,davis_2019_aa}.  

Convection has been studied somewhat extensively in non-rotating 2D/3D pre-supernova (SN) models for over the past
decade \citep{arnett_2010,arnett_2011_ab,muller_2016_aa,yadav_2019_aa,yoshida_2021_aa}.  \citet{chatzopoulos_2016_aa}
considered convection in rotating burning shells in 2D for a 20 $M_\odot$ star to find that 
rotation can influence the total power in solenoidal modes with a larger impact of C- and O-shell regions.
Recently,  3D models including rotation in the convective O-shell regions have been presented 
\citep{mcneill_2021_aa,yoshida_2021_ab}.
However, despite recent efforts, these works made approximations that present a 
challenge in accurately capturing the transport of AM near the end of the stars life. 
In the work of \citet{mcneill_2021_aa} they excised the iron core and part of the Si-shell and the simulation of
\citet{yoshida_2021_ab} was only able to capture the final $\sim$ 92 seconds prior to collapse.  The consequences of 
these approximations can affect our understanding of the AM transport within the core and potentially alter the final outcome of the star.

The AM distribution in
the stellar interior is an important aspect in predicting the properties of the compact remnant \citep{ma_2019_aa}. 
Work by \citet{summa_2018_aa} also show that the initial pre-SN AM profile,
specifically the structure of the Si/Si-O interface, can lead to \emph{qualitative} differences i.e. (explosion vs. implosion) 
in 3D rotationally-supported core-collapse supernova (CCSN) explosion models. The differences arising from 
such uncertainties also have a direct impact on the mass distribution of compact objects relevant to 
current and next-generation gravitational wave detectors (aLIGO,  VIRGO) \citep{gossan_2016_aa,pan_2018_aa,kris_2020_aa}.

In this Letter, we present a first exploratory 3D hydrodynamical simulation of the collapse of rapidly 
rotating 16 \msun star to follow the evolution of the AM of the iron core and surrounding convective burning shells. 
We directly compare our 3D simulation to predictions made by an equivalent 1D \texttt{MESA} model.  
We show that the angular momentum profile in convective regions can differ significantly between 
1D and 3D due to the efficiency of turbulent AM transport.  We find that despite the difference in the AM 
profile, the neutron star spin period estimate between the two models can agree to less than 5\%.  
For different progenitors however, 
such as those with tightly coupled Si/O convective shells,  inferred properties of the compact remnant, 
specifically the natal remnant spin rate, could disagree with \texttt{MESA} significantly.

This paper is organized as follows: in 
\S~{\ref{sec:methods}} we discuss our computational method and initial model, 
\S~{\ref{sec:results}} we present the results of our 3D simulation, 
and in 
\S~{\ref{sec:summary}} we summarize our main results.

\section{Computational Methods and Initial Model} 
\label{sec:methods}
We employ the \texttt{FLASH} simulation framework for our 3D stellar hydrodynamic model and 
utilize the  \texttt{Spark} hydrodynamics solver \citep{fryxell_2000_aa,couch_2021_aa}. 
The \texttt{Spark} hydrodynamic solver uses the weighted essentially non-oscillatory
method \citep{weno} fifth-order solver (\texttt{WENO-5}) for spatial reconstruction, a second-order 
strong stability preserving Runge-Kutta time integrator and an HLLC approximate Riemann solver. 

We assume a spherically-symmetric 
(including only the monopole term $\ell=0$) self gravitataing potential 
which includes post-Newtonian corrections
 \citep{marek_2009_aa}. We use the 21 isotope approximate nuclear reaction network implemented 
in \texttt{FLASH} \citep{timmes_2000_aa,couch_2015_aa} 
which includes updated tables for 
weak reactions \citep{langanke_2000_aa}. Our 3D simulation also uses the Helmholtz 
equation of state for a stellar plasma that 
has been used extensively \citep{timmes_2000_ab}.

Our 3D \texttt{FLASH} hydrodynamic simulation does not 
artificially enhance weak rates,  covers the full 4-$\pi$ solid angle of the star out to a spatial extent of 100,000 km 
along an arbitrary Cartesian axis and utilizes the same nuclear reaction network for the entire domain - i.e.  we do 
not excise the iron core or map it to the \texttt{MESA} model.  We use an adaptive mesh refinement 
grid with 8 levels of refinement.  The finest grid spacing is $\Delta x=$24 km out to 
a radius of $r \approx$ 2400 km (effective angular resolution of $\delta r / r \approx 1.6^{\circ}$), this radius encompasses the 
entire iron core and Si-burning shell. 

As our \texttt{FLASH} input we evolve a $m_{\rm{ZAMS}} =16~$\msun solar metallicity ($Z=0.02$) stellar model using the Modules for 
Stellar Evolution Toolkit (\texttt{MESA-r15140}) \citep{paxton_2011_aa,paxton_2013_aa,paxton_2015_aa,paxton_2018_aa}. 
This model is initialized with an equatorial velocity of 350 km / s at the surface at a point near the zero age main-sequence. 
We include different mixing processes and efficiencies following the values used in \citet{farmer_2016_aa} as well as magnetic 
torques via the ST dynamo \citep{heger_2005_aa}.
The stellar model is evolved from the pre main-sequence to a time approximately 725 seconds prior to iron core-collapse 
as determined by \texttt{MESA} as the point in 
time at which any region of the iron core experiences an infall velocity $\geq$ 1000 km / s.  

\begin{figure}[!t]
\includegraphics[width=\columnwidth]{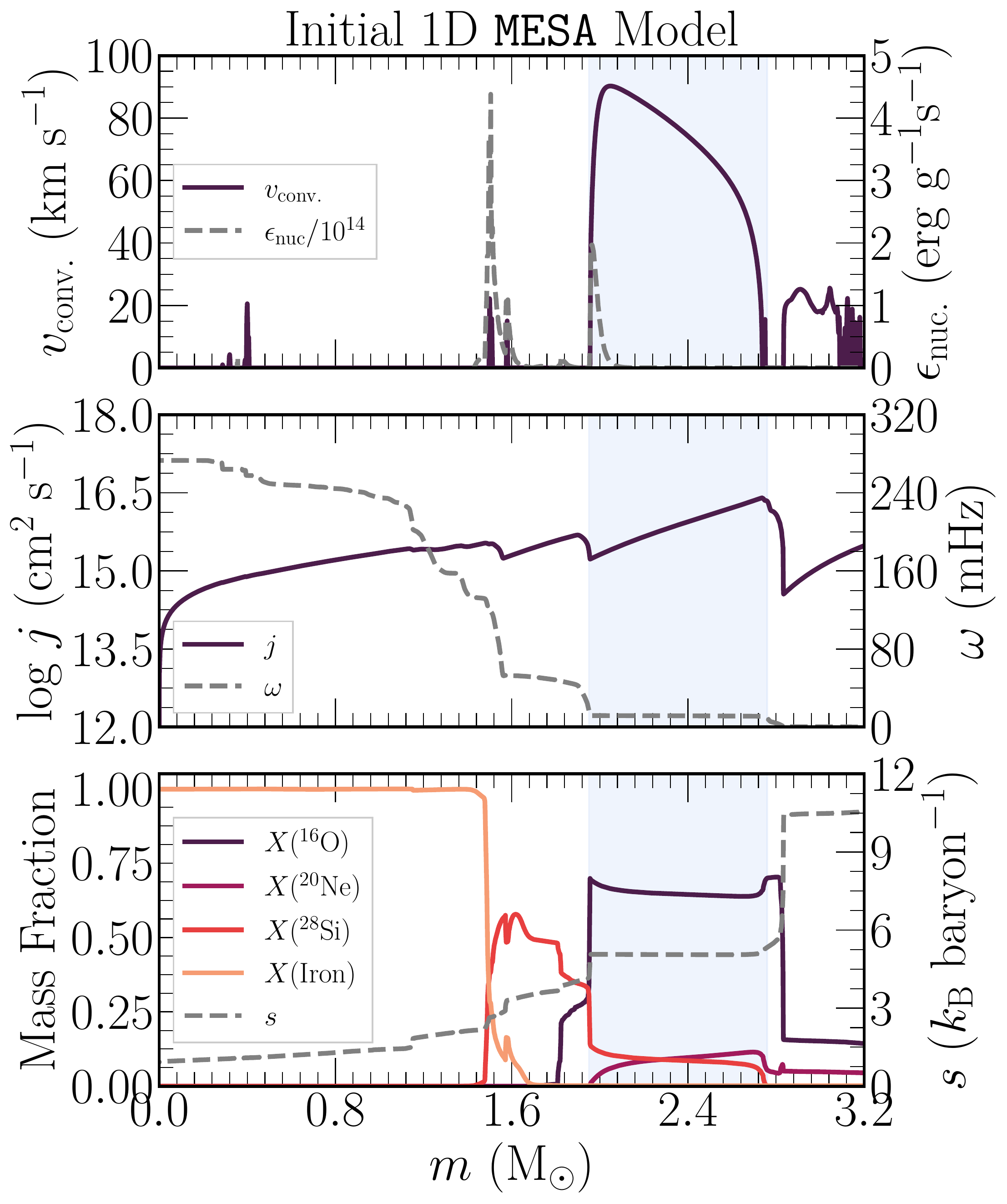}
\caption{Stellar structure profiles for the 16~\msun \MESA model at the time of mapping into \FLASH. Top panel: 
convective velocity 
according to \texttt{MLT}
and specific nuclear energy generation rate. as a function of mass coordinate.
Middle panel: specific angular momentum and angular velocity
Bottom panel: mass fraction profiles 
and the specific entropy. The convective O-shell region
is shown by the shaded blue region in each panel. 
}\label{fig:1d_structure}
\end{figure}

In Figure~\ref{fig:1d_structure} we plot various stellar structure properties of the initial 1D \MESA model at the time of mapping 
into \FLASH.  
In the top panel, we show the convective velocity according to mixing length theory (\texttt{MLT}, for an $\alpha_{\rm{MLT}}=1.6$ used 
throughout all convective regions in our 1D \MESA model) and the specific nuclear energy generation rate.   
The convective O-shell region at mapping is shown in each panel denoted by a blue shaded region 
(spanning from $m\approx 1.95-2.76$ \msun ).We see that in this region MLT predicts a peak convective 
speed of $v_{\rm{conv.}}\approx$ 90 km / s at mapping. We show the specific angular momentum and the angular 
velocity in the middle panel.  At the time of mapping the angular velocity shows speeds of $\omega \approx$ 260 mHz in the 
iron core and  $\omega \approx$ 12 mHz in the O-shell region. The rotational and convective velocity speeds in the O-shell 
region are comparable at this point in time.  We can compute the 
convective turnover timescale in this region as $t_{\rm{conv.,O}} \approx \Delta r_{\rm{O}} /v_{\rm{conv.,O}}\approx159$ s.
In the bottom panel of Figure~\ref{fig:1d_structure} we show the 
mass fraction profiles of our input \MESA model 
showing the most dominant isotopes. Beneath the convective O-shell region lies a narrow Si-shell burning region. 
Our model shows an iron core mass of $M_{\rm{Fe}}\approx1.46$ \msun at the time of mapping.

\begin{figure}[!t]
\centering{\includegraphics[width=\columnwidth]{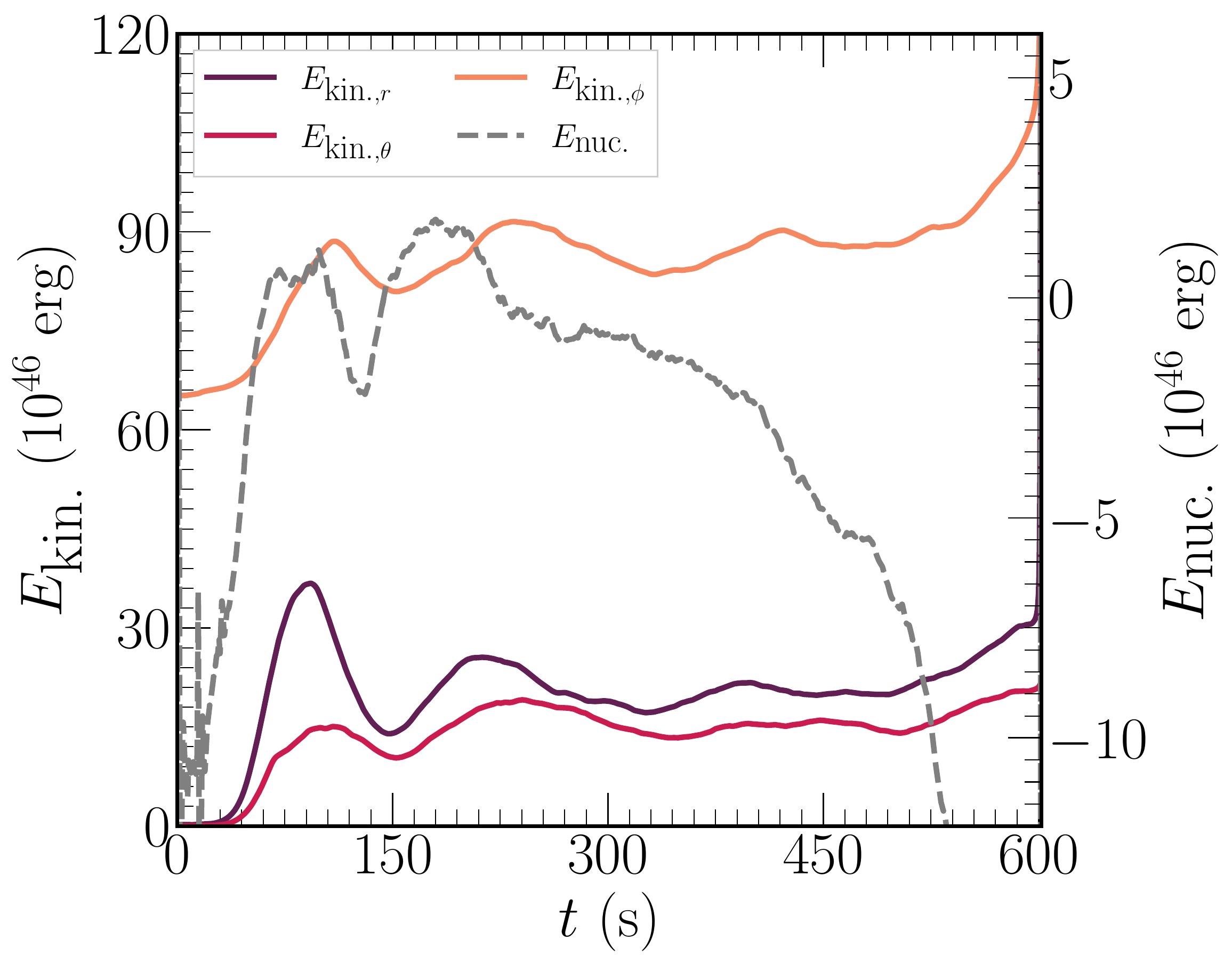}}
\caption{Time evolution of the integrated kinetic (left) and nuclear (right) energy for the duration of the 3D \FLASH simulation. }
\label{fig:3d_E_kin}
\end{figure}

We follow the methods described in \citet{zingale_2002_aa} for mapping the 1D initial model into our 3D \texttt{FLASH} domain. 
To initiate convection in the O-shell we use the methods described in \citet{oconnor_2018_aa} in which we choose a 
scaling factor and spherical harmonic degree $\ell$ to apply perturbations to the velocity field.  We choose an $\ell$ of 
9 and a scaling factor $C$ such that the angle-average velocity reproduces 5\% of the convective velocity speed as 
predicted by \MESA at mapping.  Because of the narrow Si-shell burning region and the low convective speeds we do not apply 
initial perturbations to the Si-shell region.  Constant angular velocity as a function of radial coordinate (shellular rotation) is
assumed initially in the 3D \texttt{FLASH} simulation using 
the 1D profile as provided by \texttt{MESA}. In \texttt{FLASH}, rotation is implemented to be about the $y-$axis.

\section{3D Collapse of a Rotating 16 \msun Star}
\label{sec:results}
We follow the 3D hydrodynamic evolution up to and including gravitational instability and iron core-collapse in our 
\texttt{FLASH} simulation.  The simulation is evolved for a total of $t\approx$ 602 s at which point the core has began 
to experience iron core-collapse and has reached an infall velocity of 1595 km / s.  
In Figure~\ref{fig:3d_E_kin} we show the integrated kinetic and nuclear energy for the 
3D \FLASH simulation. 
We note an initial transient phase in the radial kinetic energy that 
shows a local maxima at $t\approx100$ s. Following this time, the 3D simulation shows a quasi-steady state 
represented by a balance between driving nuclear energy generation and the kinetic energy in the convective O-shell region. 
At a time beyond $t\geq$ 500 s contraction of the iron core begins to accelerate leading to an overall increase in the 
total kinetic energy and overall decrease of the net nuclear energy due to an increase in the negative specific energy generation 
rate in the iron core.  

\begin{figure}[!t]
\centering{\includegraphics[width=\columnwidth]{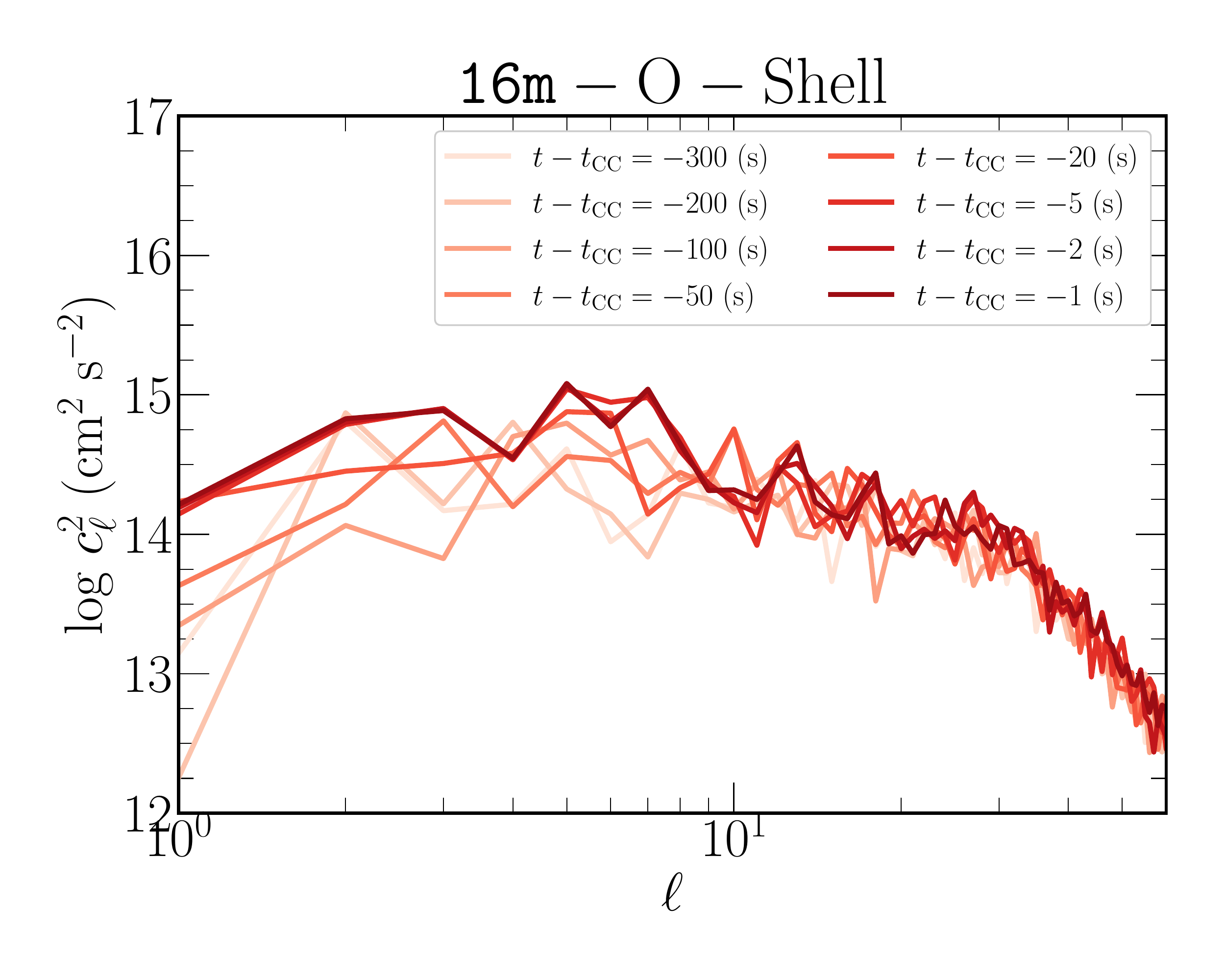}}
\caption{Time evolution of the power spectrum for the spherical harmonic decomposition of the turbulent 
radial velocity field for the convective O-shell region. The spectra are evaluated in the 3D model at eight different 
times at a radius of $r_{\rm{O-shell}}$ = 6000 km.}
\label{fig:3d_o_shell_spectra}
\end{figure}

\begin{figure*}[!t]
\centering{
\includegraphics[width=2.1\columnwidth]{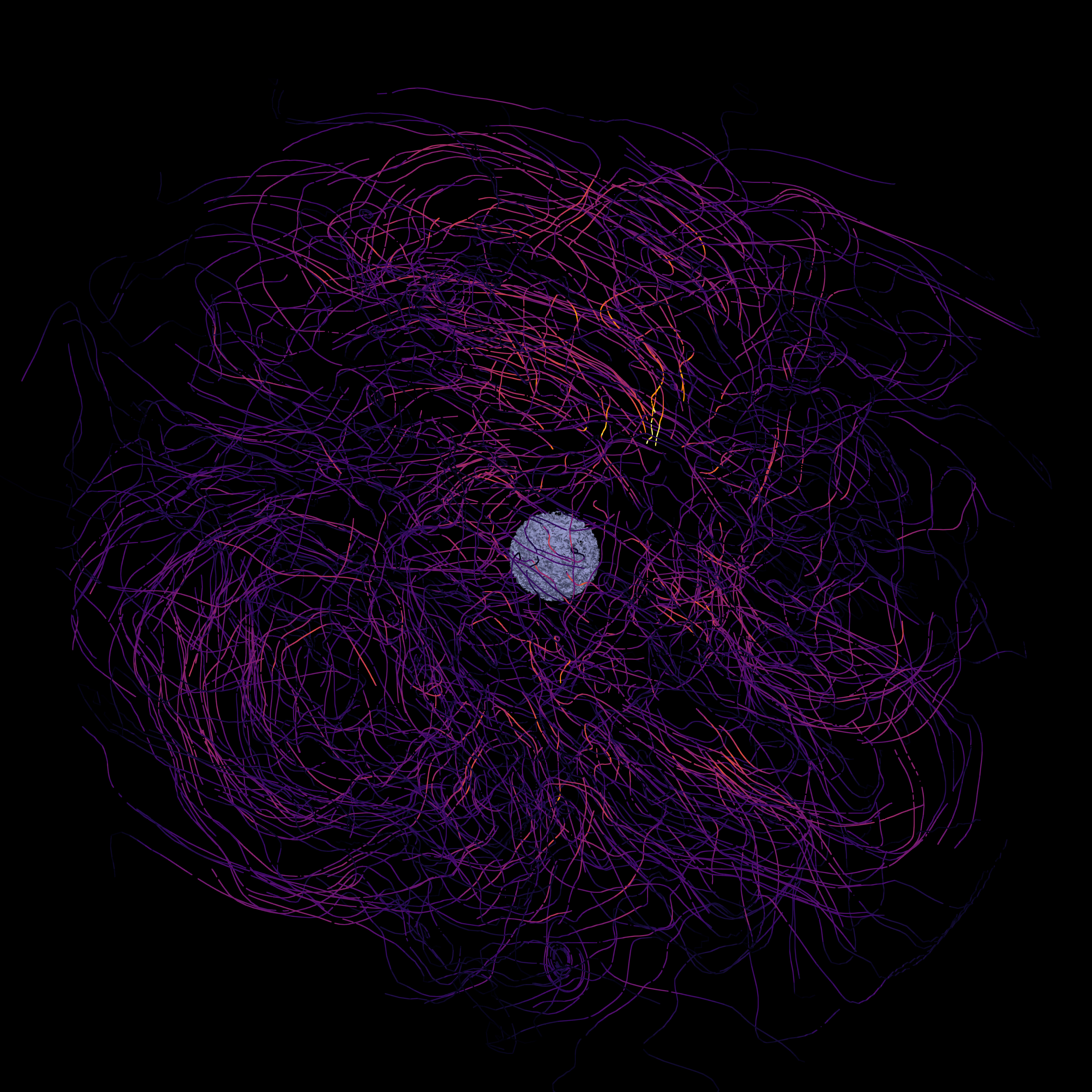}
}
\caption{Streamlines of the magnitude of the velocity field for the 3D \texttt{FLASH} simulation at a time $\approx$ 7 seconds
before iron core-collapse. The purple contour shows the iron core at a radius of $r\approx 2400$ km shown here. The brighter (orange) stream 
lines correspond to speeds of $\left | \textbf{v} \right |\approx 800 $ km / s. The orientation of the rendering is shown with the north 
vector aligned with the $y$-axis, the axis of rotation, (with a slight horizontal offset).  Large scale flow is observed with 
streamlines wrapping about the axis of rotation out to the edge of the convective shell region. The spatial scale of the image 
corresponds to a spatial extent of $\approx$ 40,000 km on a side. This image was produced using \texttt{Paraview}.
}
\label{fig:3d_omega}
\end{figure*}

\subsection{Characterizing the convection and flow morphology}
To characterize the scale of the convective eddies, we decompose the fluctuating 
component of the radial velocity field into spherical harmonics \citep{sht_2013_aa}. The total power in a given 
mode is given by 
\begin{equation}
c^{2}_{\ell} = \sum^{\ell}_{-\ell} \left | \int Y^{m}_{\ell} (\theta,\phi) 
v^{\prime}_{\rm{rad.}} (r_{\rm{O-shell}},\theta,\phi) d \Omega \right | ^{2}~,
\end{equation}
where $Y^{m}_{\ell}$ is a given spherical harmonic of degree $\ell$ and order $m$.  The 
turbulent radial velocity $v^{\prime}_{\rm{rad.}}$ is computed as 
$v^{\prime}_{\rm{rad.}} = v_{\rm{rad.}} - \left < v_{\rm{rad.}} \right >$, where the closed brackets 
denote an angle-average of a given quantity.
For the convective O-shell region we choose a value of $r_{\rm{O-shell}}$ = 6000 km.

In Figure~\ref{fig:3d_o_shell_spectra} we show the time evolution of the power of the 
spherical harmonic decomposition for the turbulent radial velocity field for eight different times in the 
3D \texttt{FLASH} model. At $t-t_{\rm{CC}}=-300$ s,  the spectra show a peak at $\ell=2$ suggesting the 
bulk of the radial kinetic energy is stored at the largest scales. For the last 300 seconds, a steady increase in power is 
observed in the lowest modes ($\ell=1-3$).  As the model approaches collapse, the peak shifts 
to $\ell=5$ at approximately two seconds prior to collapse.  The dominant modes containing the most
power are $\ell=2,3,5$ and $\ell=7$ one second before collapse.  
In Figure~\ref{fig:3d_omega} we show a streamlines of the magnitude of the velocity field depicting the
qualitative structure of the convective O-shell region at $t\approx595.5$ s,  $\approx$ 7 seconds prior to core-collapse.

The Solberg-H\o{}iland (SH) instability can arise in regions of significant rotational shear.  This instability can 
work to stabilize convection in a region that would otherwise be unstable due to the Ledoux 
criterion \citep{stothers_1973_aa,pajkos_2019_aa}.
The contribution of this gradient would be considered as an additional term 
to the \BV~frequency $N^{2}$ \citep{endal_1978_aa,heger_2000_aa},
\begin{equation}
N^{2}_{\Omega} = r^{-3} \frac{d}{dr}\left (j^2_{y} \right ) \geq 0~.
\end{equation}
In our 3D simulation we find that the magnitude of the shear is not sufficient to significantly alter the 
convection in the O-shell region and that near collapse the specific angular momentum gradient tends towards 
a uniform value thus moving further from a region where the SH would be relevant.  

Differential rotation was found to operate in 3D simulations of O-shell burning \citep{mcneill_2021_aa}. In their model, 
they observed a Rossby number of $\approx0.24-0.4$ during the quasi-steady convection state.  We can define the 
convective Rossby number as 
\begin{equation}
\textup{Ro}  = \frac{ \bar{v}_{\rm{conv.}} }{ 2  \bar{\omega} \Delta r }~,
\end{equation}
where the bar denotes averages over the convective O-shell region and $\Delta r$ is the width of the O-shell.
In our simulation we observe a value of Ro$\approx$0.15 at a time $t\approx420$ s.

\subsection{Angular Momentum Transport}
To begin, we compute the integrated $y-$component of angular 
momentum contained in our 3D domain at $t=0$ and compare this to the point at collapse to assess the AM conservation 
throughout our simulation. In doing so, we find an initial value of $J_{y}(t=0)\approx 2.76\times10^{49}$ g cm$^{2}$ s$^{-1}$. 
At a time of $t\approx200$ s,  we compute a total of $J_{y}(t=200)\approx 2.74\times10^{49}$ g cm$^{2}$ s$^{-1}$
and at the end of the simulation, 
$J_{y}(t=t_{\rm{CC}})\approx 2.70\times10^{49}$ g cm$^{2}$ s$^{-1}$. This result suggests a loss of 2.4\% of the total 
$y-$ component of angular momentum over the duration of the simulation.
\citet{mcneill_2021_aa} estimate angular momentum conservation of $\approx$ 11 \% for their rotating 3D O-shell simulation.

\begin{figure*}[!t]
\centering{\includegraphics[width=\columnwidth]{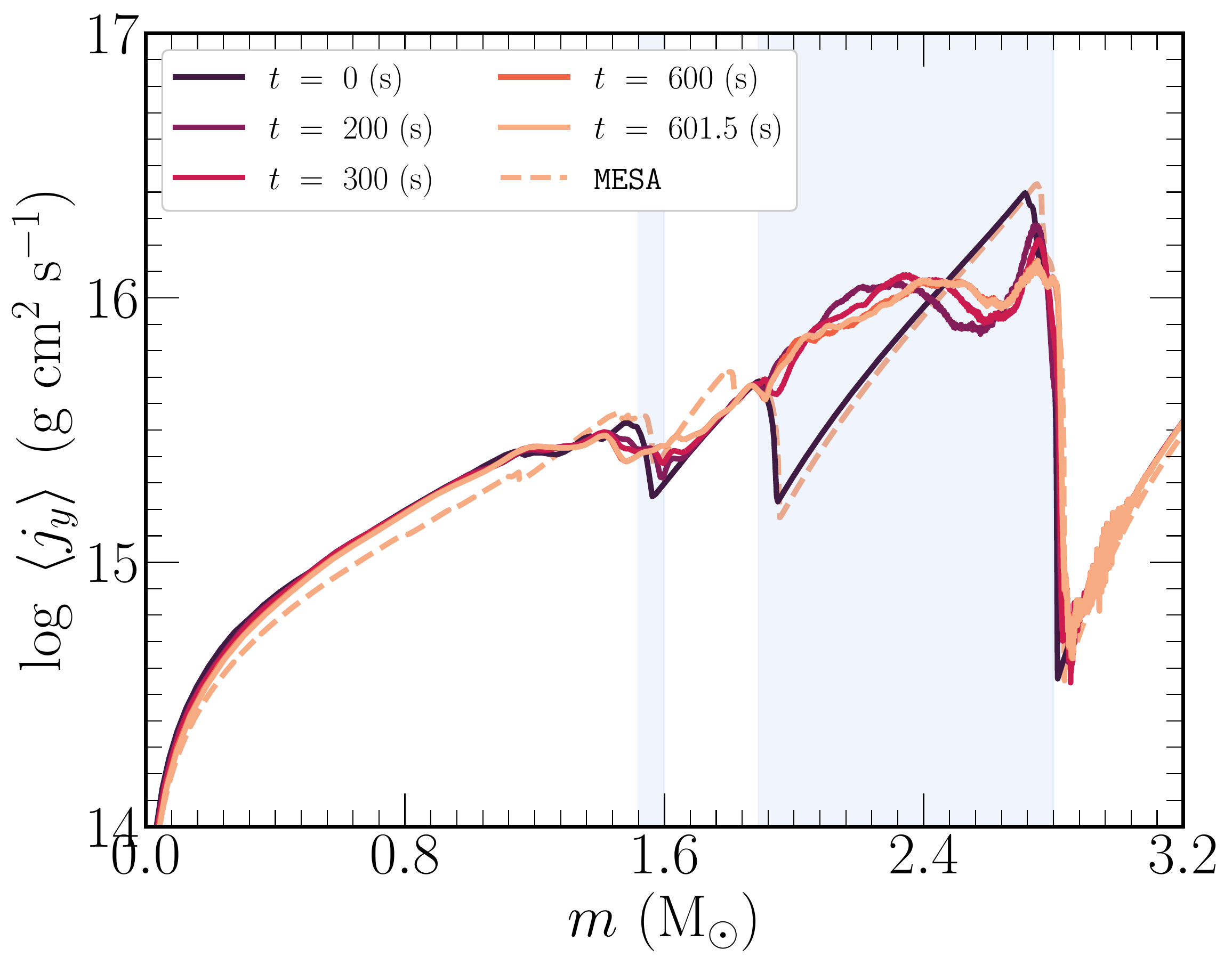}}
\centering{\includegraphics[width=\columnwidth]{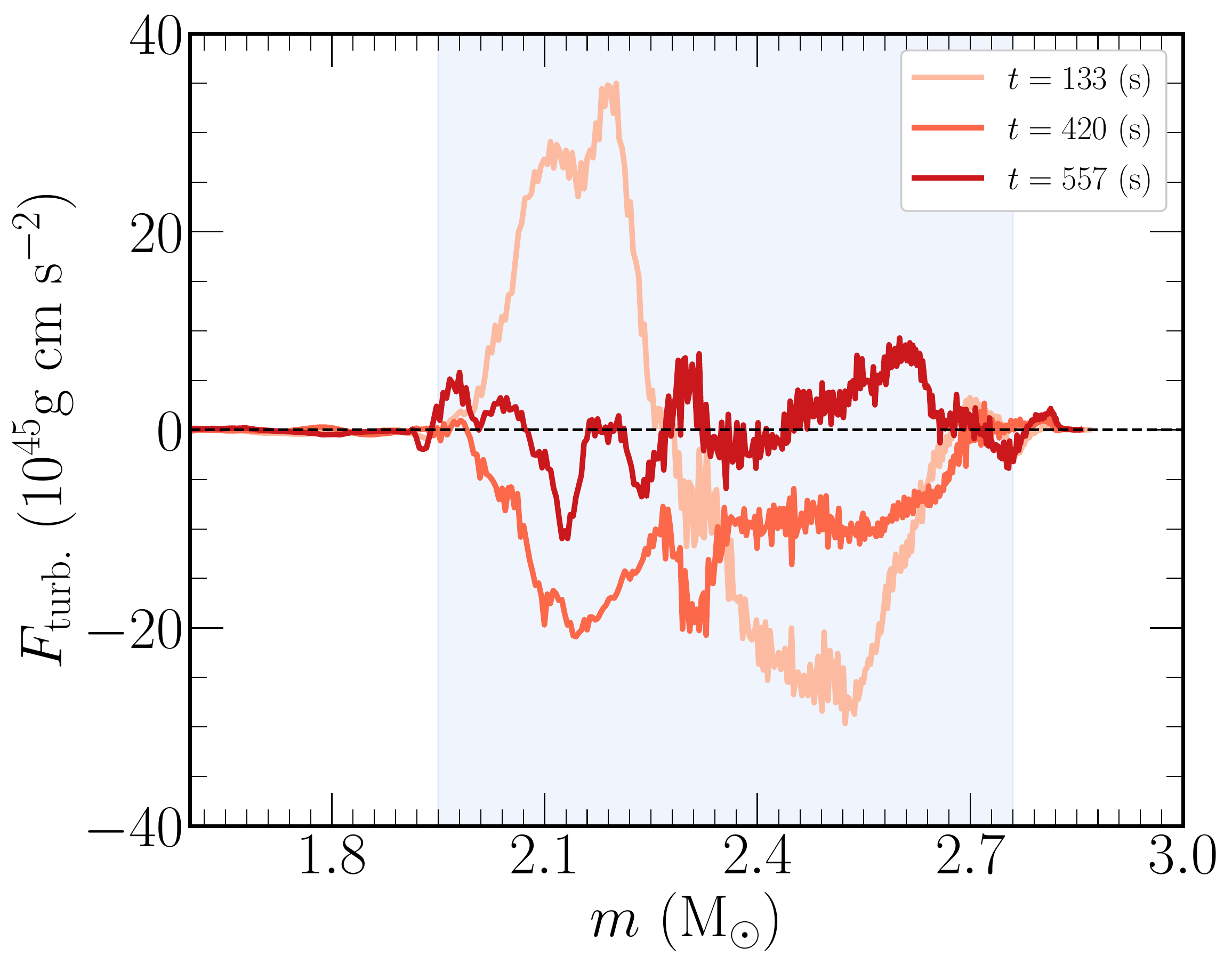}}
\centering{\includegraphics[width=\columnwidth]{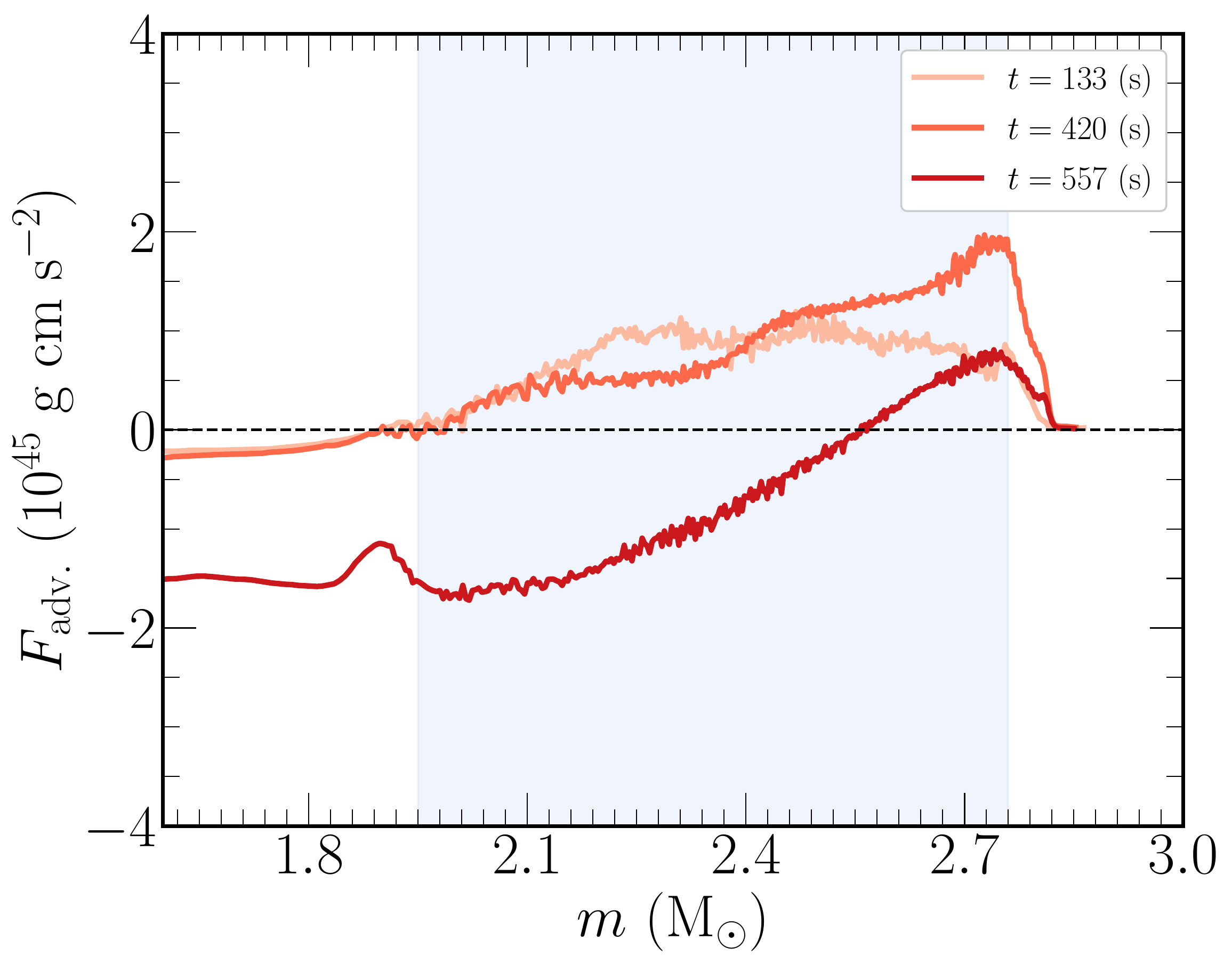}}
\caption{\emph{Top left:} Angle-average of the $y-$ component of the specific angular momentum for the 3D \texttt{FLASH} simulation 
at five different times include $t=601.5$s (half a second before iron core-collapse). The correponding \texttt{MESA} model 
is also shown the dashed gray line chosen to match the central density of the 3D model at $t=601.5$s . 
Turbulent ($4\pi r^2 F_{\rm{turb.}}$, \emph{Top right}) and advective flux ($4\pi r^2 F_{\rm{adv.}}$, \emph{Bottom left}) 
terms for the 3D simulation at 
three different times. 
The approximate convective Si-($m\approx1.52-1.60 M_{\odot}$) and O-shell ($m\approx1.89-2.80 M_{\odot}$) regions 
at $t=601.5$ are denoted by the blue shading. }
\label{fig:3d_j_rot}
\end{figure*}

In Figure~\ref{fig:3d_j_rot} we show the time evolution of the angle-averaged $y$-component of the specific angular 
momentum ($j_{y}$) from our 3D simulations at five different times during the simulation.  The approximate extent of the convective 
Si- and O-shell region is denoted by the blue shading. Because the 3D simulation collapses $\approx$ 123 \emph{earlier}
than predicted by \texttt{MESA}, we compare our 3D profiles at collapse to a \texttt{MESA} 1D profile that most closely matches the
central density at the point of comparison.  The \texttt{MESA} models used for comparison was also started from the 
point of mapping into \texttt{FLASH} and evolved to CC \emph{without} the ST dynamo as magnetic fields are 
not considered in the 3D simulation presented here.

In general, we find that AM profile in the iron core (inner $m\lesssim1.6$\msun) remains constant in shape 
only showing a slight decrease in the core from $t=0$ to the point of collapse.  The \texttt{MESA} model shows a decrease in 
angular momentum over this time as angular momentum appears to be diffused out to the outer Si-shell region.  
At a mass coordinate range corresponding to the Si-shell region, $m\approx1.6-1.95$\msun we notice two things in the 3D 
angle-average AM profiles. First, the steep gradient at $t=0$ is smoothed out and tends towards a uniform 
profile near the active and at some points convective Si-shell burning region ($m\approx1.6-1.7$\msun). Secondly,  beyond this 
mass coordinate ($m\approx1.7-1.95$\msun), the 3D model is not convective and the AM profile does not vary significantly 
over time. The AM profile 
matches closely for the 3D profile at $t=0$ to that of the 1D  \texttt{MESA} profile but shows less AM in this region at collapse. 

We observe a similar feature in the convective O-shell region, $m\approx1.95-2.76$\msun where the steep 
angular momentum gradient is smoothed efficiently (as quickly at $t=200$ s) and the \emph{sign} of the angular momentum 
gradient at the base of the O-shell is reversed and maintained until collapse.  Throughout most of the convective O-shell 
region the profile
tends towards a uniform 
profile near $\textup{log} \left < j_{y} \right > \approx 16$ g cm$^{2}$ s$^{-1}$.  Beyond the convective O-shell, the 
profile shape matches well to the initial configuration given by \texttt{MESA} as no active convection is occurring in the 
3D simulation or 1D  \texttt{MESA} model. 
 
One can determine the relevant contribution of angular momentum transport in the convective O-shell region by 
considering the Favre-average angular momentum transport equations.
Here, we follow the methods described in \citet{mcneill_2021_aa} and whose derivation is described in detail 
in \citet{mcneill_2020_aa}.  In this study we only present the two relevant terms for angular momentum transport in the 
O-shell,
\begin{equation}
\frac{\partial \left < \rho \textbf{j}_{y} \right >}{\partial t} + \nabla \cdot \textbf{F}_{\rm{adv.}} + \nabla \cdot \textbf{F}_{\rm{turb.}}=0~.
\end{equation}
The advective flux can be computed as
\begin{equation}
F_{\rm{adv.}} = \rho \tilde{v}_{\rm{r}} \tilde{j}_{y}~,
\end{equation}
where $v_{\rm{rad.}}$ is again the radial velocity and $j_{y}$ is the $y-$component of the 
specific angular momentum magnitude.  
Here, the tilde defines a Favre-averaged quantity \citep{favre_1965_aa} 
$\tilde{X}$ as
\begin{equation}
\tilde{X} = \frac{\int \rho X \textup{d} \Omega}{\int \rho \textup{d} \Omega}~.
\end{equation}
The turbulent flux component can be computed as,
\begin{equation}
F_{\rm{turb.}} = \left < \rho v^{\prime \prime}_{\rm{r}} j^{\prime \prime} \right > ~.
\end{equation}
We adopt the same notation given in \citet{mcneill_2020_aa} where $X^{\prime \prime} = X - \tilde{X}$.

In Figure~\ref{fig:3d_j_rot} we plot the resulting advective (\emph{bottom}) and turbulent (\emph{top right}) 
fluxes at $t=113$ s (once convection 
has been established), $t=420$ s (towards the end of the quasi-steady convective state in the O-shell) and $t=557$ s 
(as the model approaches collapse and evolves out of dynamical equilibrium). At the
earliest time considered we observe a positive flux (positive $v^{\prime \prime}_{\rm{rad.}}$) 
in the O-shell at $m\approx1.9-2.3$ \msun that peaks at 
$\approx 3\times10^{46}$ g cm s$^{-2}$, 
from $m\approx$ 2.3-2.8  \msun the same magnitude is found but for a negative flux.  At this point in time, the turbulent component
is causing the observed change in the AM profile throughout the convective O-shell observed in Figure~\ref{fig:3d_j_rot}.

For $t=420$ s,  we observe a negative
flux across the entire convective O-shell region with a peak value of $\approx 2\times10^{46}$ g cm s$^{-2}$.  At this time, the 
AM profile appears to have reached a new equilibrium in the redistribution with a modest net flux inward. 
For the final 
time considered, the magnitude of the flux decreases to about $\approx 1\times10^{46}$ g cm s$^{-2}$ suggesting turbulent 
transport has become less efficient except for at a mass coordinate of $m \approx 2.4-2.7$ \msun where a significant AM gradient
still exists but turbulent radial velocity speeds are lower and require a longer time to homogenize the AM profile. 

The advective flux component (bottom panel of Figure~\ref{fig:3d_j_rot}) shows positive flux 
for times $t=133$ s and $t=420$ s across the entire O-shell region with a peak value near the edge of the 
O-shell of  $4\pi r^2 F_{\rm{adv.}}$ $\approx 2\times10^{45}$ g cm s$^{-2}$.  These values suggest advection of 
AM out of the convective O-shell region also observed via the gradual decrease over time of the peak of the specific 
AM at the outer convective boundary in Figure~\ref{fig:3d_j_rot}. At $t=557$ s,  a negative flux is found 
with magnitude of $\approx 2\times10^{45}$ g cm s$^{-2}$ over most of the O-shell. The shift towards a negative 
flux for this term is likely attributed to the acceleration of the contraction leading to an increase of the inward (negative)
radial velocity in the O-shell. 

At late times ($t\lesssim$ 300 s before collapse), the Si-shell is convective over a thin region. 
Within this region, a positive turbulent flux is observed which helps homogenize the
angular momentum profile in a similar fashion to that of the O-shell. 
However, the turbulent flux component in this region is found to be an order of magnitude less than those in the O-shell due to
lower turbulent radial velocity speeds and leads to a longer time needed 
for the angular momentum profile to readjust. For example in Figure~\ref{fig:3d_j_rot} at $t=200$ s the AM 
profile in O-shell region is qualitatively different and the gradient at the convective boundary has changed sign. 
However,  the non-convective Si-shell region ($m\approx1.6-1.95$\msun) still shows partial correspondence 
to the shape of the 
 \texttt{MESA} AM profile up to the point of collapse. 
Near the convective Si-shell burning region ($m\approx1.5-1.6$\msun), the AM profile begins 
to deviate from the 1D \texttt{MESA} profile moving closer towards uniform in the evolution towards collapse.  

3D models of rotationally-supported CCSN explosions were shown to lead to qualitative differences when considering
different angular momentum profiles \citep{summa_2018_aa}. In their comprehensive study, the found that only their 
"artifical rotation" model which utilized an enhanced AM profile was able to reach the conditions for shock revival and 
explosion as compared to an AM profile from a 1D stellar evolution model. The determining factor was found to be 
contributed to the time at which the Si/Si-O interface was accreted onto 
the shock. The results of their study suggest a crucial dependence on the specific AM profile in the Si-/O-shell 
convective regions near collapse.  The efficient transport of AM facilitated by convection in the 3D model
considered here is a first step in understanding the impact such progenitor models might have on models of CCSNe. 

\subsection{Proto-Neutron Star Mass and Spin Estimates}
The AM profile predicted from stellar evolution models can be used to determine properties of the 
proto-neutron star (PNS) or black hole (BH) borne as a result of stellar core-collapse. 
Following \citet{heger_2005_aa}, we compute inferred PNS properties using the 3D rotating model 
presented here as well as the 1D equivalent \texttt{MESA} model. 
We assume a 12 km neutron star and a moment of inertia of the form 
$I\approx 0.35 M_{\rm{grav}} R^{2}$ for the estimates presented here.  For both models we 
find $M_{\rm{grav.}}\approx1.58$ \msun 
and $M_{\rm{bary.}}\approx1.88$ \msun for the gravitational and baryonic 
mass values, respectively.  
The resulting period for the 3D model at $t=602$ s is found to be 
1.57 ms, agreeing with the estimate for the \texttt{MESA} model to within $\approx$ 1\%. 
When using a fixed mass cut of $M_{\rm{grav.}}\approx1.4$ \msun and  $M_{\rm{bary.}}\approx1.7$ \msun
for both models, a larger difference is observed with \texttt{MESA} predicting a $\approx$ 5\% longer period.  

While we find agreement with \MESA in the natal PNS spin rate for the 3D progenitor model considered here,  models 
with different Si/O convective structures might lead to larger differences in the integrated angular momentum profile.
Such profiles are crucial in predicting natal BH and PNS spin rates and also sensitive to input physics in 
1D stellar models. The combination of these uncertainties present a challenge in our predictive capability in describing 
the mass distribution and spin rates of binary NS and binary BH systems observed by Advanced LIGO \citep{kris_2020_aa}. 
Rotating 3D CCSN progenitor models such as the exploratory model presented here 
represent a necessary step towards addressing these challenges.

\section{Summary and Discussion}
\label{sec:summary}
We have presented the evolution of the AM distribution 
in the final 10 minutes before iron core-collapse in a rapidly rotating 16~\msun star revealed by a $4\pi$ 3D 
hydrodynamic simulation. For the first time, we were able to directly compare AM profile in the stellar core and 
surrounding convective shells in our  
3D simulation to a 1D model predicted by \texttt{MESA}.  It was shown that despite the different 
dynamical evolution of the stellar core in the 3D simulation the 
proto-neutron star spin period estimate at core-collapse can agree to \texttt{MESA} less than 5\%. 
For different progenitor models however, 
such as those with tightly coupled Si/O convective shells,  the efficient AM distribution in 3D simulations may lead to 
differences in the inferred remnant properties as compared to \texttt{MESA}.

Angular momentum transport was found to be facilitated in the convective O-shell region via a positive 
turbulent flux allowing the shape of the profile to reach a nearly uniform specific angular momentum within a single convective turnover 
time. In non-convective Si-rich regions it was shown that the profile matched well to \texttt{MESA}. Near the convective 
Si-shell region of the 3D model, the AM profile began to diverge from the shape predicted by \texttt{MESA} but on a slower
timescale due to the lower turbulent speeds when compared to the O-shell region. 

At collapse, the peak of the turbulent velocity spectrum is 
found at spherical harmonic degree $\ell=5$.  Recent work by \citet{yoshida_2021_ab} observed large-scale spiral arms 
emerge prior to collapse in the rotational kinetic energy density distribution. We did not identify 
such spiral arm structures in the 3D model presented here.
As with the presence of large-scale modes in the convective 
shells near collapse, rotation is a crucial component and must be taken into consideration in ordinary 3D CCSN progenitor 
models and those that might produce more luminous transients.  

The consequences of the AM transport observed in the 
3D model presented here could \emph{qualitatively} alter rotationally-supported CCSN explosion dynamics and inferred 
remnant properties. 
The exploratory 3D model presented here is a first step in considering the effects of rotation within the stellar core and 
surrounding convective shells.  Long-term, high-resolution models informed by a range of initial 1D models 
will be necessary to determine the complex interplay of turbulence, rotation, and convection in the fate of a massive 
star. 

\software{
\MESA \citep[][\url{http://mesa.sourceforge.net}]{paxton_2011_aa,paxton_2013_aa,paxton_2015_aa,paxton_2018_aa},
\FLASH \citep[][\url{http://flash.uchicago.edu/site/}]{fryxell_2000_aa},
\texttt{yt} \citep[][\url{https://yt-project.org}]{turk_2011_aa}, and
\texttt{matplotlib} \citep[][\url{https://matplotlib.org}]{hunter_2007_aa}.}

\acknowledgements
The author thanks
Sean Couch,
Josh Dolence,
and 
Philipp Edelmann
for useful discussions. 
Research presented in this article was supported by the 
Laboratory Directed Research and Development program of Los Alamos National Laboratory under 
project number 20210808PRD1. 
This work was supported in part by Michigan State University through 
computational resources provided by the Institute for Cyber-Enabled Research.
This research made extensive use of the SAO/NASA Astrophysics Data System (ADS).

\bibliographystyle{aasjournal}
\bibliography{prog3dn}

\listofchanges

\end{document}